\documentclass[12pt]{article}
\usepackage{epsf}
\usepackage{graphicx}
\usepackage{array,dcolumn}      
\usepackage{axodraw}
\usepackage{amssymb}
\newcommand{\hs}{\hspace*{0.5cm}}
\newcommand{\be}{\begin{equation}}
\newcommand{\ee}{\end{equation}}
\newcommand{\bea}{\begin{eqnarray}}
\newcommand{\eea}{\end{eqnarray}}
\newcommand{\baa}{\begin{eqnarray*}}
\newcommand{\eaa}{\end{eqnarray*}}
\newcommand{\bary}{\begin{array}}
\newcommand{\eary}{\end{array}}
\newcommand{\bit}{\begin{itemize}}
\newcommand{\eit}{\end{itemize}}

\begin{document}
\begin{center}
{\large \bf    BILEPTON  PRODUCTION
IN $ e^-\  \gamma $ COLLISIONS} \vspace*{1cm}

{\bf $ \mbox{Dang Van  Soa}^{a,}
\footnote{On leave of absence from Department of Physics,
Hanoi University of Education, Hanoi, Vietnam}$},
{\bf$ \mbox{Takeo Inami}^a, 
\mbox{Hoang Ngoc Long}^{b,}
\footnote{On leave of absence from Institute of Physics, NCST, 
P.O. Box 429, Bo Ho, Hanoi 10000,
Vietnam} $}\\[0.2cm]
{\it $^a$ Department of Physics,
Chuo University, Kasuga 1-13-27, Bunkyo, Tokyo 112, Japan}\\
{\it $^b$ The Abdus Salam International Centre for Theoretical
Physics, Trieste, Italy }\\
\end{center}
\vspace*{1cm}

\begin{abstract}

We study the production of bileptons, new gauge boson of lepton
number two, in the minimal 3 - 3 - 1 model in high energy $e^-
\gamma$ collisions. If the bilepton masses are in the range of 300
GeV the reaction will give observable cross-sections in the future
colliders.

\end{abstract}
\vspace*{0.5cm}

PACS number(s):12.10.-g, 12.60.-i, 13.10.+q, 14.80.-j.\\


\section{Introduction}

\hs All models of physics beyond the standard model (SM) predict
the existence of new particles. One of the more exotic of these is
bilepton, a particle of lepton number 2. Bileptons occur in a
few of models of new physics. One of them is the gauge models
based on the
 $\mbox{SU(3)}_C \otimes \mbox{SU(3)}_L
\otimes \mbox{U(1)}_N $ (3 - 3 - 1) gauge group [1-5].
These models contain a number of intriguing features: First, the
models predict three families of quarks and leptons if the anomaly
free condition on $ \mbox{SU(3)}_L \otimes \mbox{U(1)}_N$ and the
QCD asymptotic freedom are imposed. Second, the Peccei-Quinn
symmetry naturally occurs in these models~\cite{pal}. The third
interesting point is that one generation of quarks is treated
differently from two others~\cite{lng} . This could lead to a natural
explanation for the unbalancingly heavy top quark.

\hs It has recently been argued that, the 3 - 3 - 1 model arises
naturally from gauge theories in spacetime with
 extra dimensions~\cite{qui}, where the scalar fields
are the components in the extra space dimensions of a
higher dimensional gauge field~\cite{hil}. A few different
versions of the 3 - 3 - 1 model have been proposed~\cite{col}.

\hs The production of doubly charged vector bileptons in high energy
collisions has been widely discussed both in the generic model~\cite{fcu}
and in the minimal model~\cite{fra}.
Recent investigations have indicated that signals of new gauge bosons
in these models may be observed at the CERN LHC~\cite{bdi} or
Next Linear Conllider (NLC)~\cite{mpt}. Production of scalar
leptoquarks~\cite{ebo} and constraints on extra gauge bosons~\cite{sgo}
in $e\gamma$ collisions were studied.
The potentiality of the projects of TESLA and CLIC based $e\gamma$ colliders
were discussed in~\cite{rwa}.

In~\cite{ls}
the trilinear gauge boson couplings were presented and production
of bileptons in $e^+  e^-$ collisions was calculated. In this
paper we turn our attention to the  future perspective experiment, namely
$e^- \gamma$ collision. Finally, the conclusions are presened in Sec. IV.
This paper is organized as follows. In Sec.II we summarize the
basic elements of the minimal 3 - 3 - 1 model. Sec.III is devoted
to single bilepton production in $\gamma \ e^-$ collisions.
Finally, the conclusions are presented in Sec. IV.

\section{A review of the minimal 3 - 3 - 1 model }

\hs To frame the context, it is appropriate to recall briefly  some
relevant features of the minimal  3 - 3 - 1 model~\cite{pp,fr}.
The model treats the leptons as $\mbox{SU(3)}_L$ antitriplets~\cite{fr,dng}
\be f_{a L} = \left( \begin{array}{c}
               e_{a L}\\ -\nu_{a L}\\  (e^c)_{a L}
               \end{array}  \right) \sim (1, \bar{3}, 0),
\label{l} \ee
where $ a = 1, 2, 3$ is the family index.\\
\hs Two of the three quark generations transform as triplets of the
$\mbox{SU(3)}_L$  and
the third is treated differently. It belongs to an
antitriplet.
 \be Q_{iL} = \left( \begin{array}{c}
                u_{iL}\\d_{iL}\\ D_{iL}\\
                \end{array}  \right) \sim (3, 3, -\frac{1}{3}),
\label{q} \ee
\[ u_{iR}\sim (3, 1, 2/3), d_{iR}\sim (3, 1, -1/3),
D_{iR}\sim (3, 1, -1/3),\ i=1,2,\] \be
 Q_{3L} = \left( \begin{array}{c}
                 d_{3L}\\ - u_{3L}\\ T_{L}
                \end{array}  \right) \sim (3, \bar{3}, 2/3),
\ee
\[ u_{3R}\sim (3, 1, 2/3), d_{3R}\sim (3, 1, -1/3), T_{R}
\sim (3, 1, 2/3).\]\\
\hs The nine gauge bosons  $W^a (a = 1, 2, ..., 8)$ and $B$ of
$\mbox{SU(3)}_L$ and $\mbox{U(1)}_N$ are split into four light 
gauge bosons and
five  heavy gauge bosons after $ \mbox{SU(3)}_L \otimes \mbox{U(1)}_N$
is broken to $\mbox{U(1)}_Q$. The light gauge bosons are those of the
standard model, $A, Z_1$ and $\ W^\pm$. The new heavy gauge bosons are
$Z_{2}$, $\ Y^\pm$ and the  doubly charged bileptons $X^{\pm \pm}$. They
are expressed  in terms of $W^a$ and $B$  as~\cite{dng}
\bea \sqrt{2}\ W^+_\mu &=& W^1_\mu - iW^2_\mu ,
\sqrt{2}\ Y^+_\mu = W^6_\mu - iW^7_\mu ,\nonumber\\
\sqrt{2}\ X_\mu^{++} &=& W^4_\mu - iW^5_\mu, \label{idmin1} 
\eea
\bea 
A_\mu  &=& s_W  W_{\mu}^3 + c_W\left(\sqrt{3}\ t_W\ W^8_{\mu}
+\sqrt{1- 3\ t^2_W}\  B_{\mu}\right),\nonumber\\
Z_\mu  &=& c_W  W_{\mu}^3 - s_W\left(\sqrt{3}\ t_W\ W^8_{\mu}
+\sqrt{1- 3\ t^2_W}\  B_{\mu}\right),\nonumber\\
Z'_\mu &=&-\sqrt{1- 3\ t^2_W}\ \ W^8_{\mu} + \sqrt{3}\ t_W\
B_{\mu}, \label{apstat}
\eea 
where we use the
notation $c_W \equiv \cos \theta_W$, $s_W \equiv \sin \theta_W $
and $t_W \equiv \tan \theta_W$. The {\it physical} states are
mixtures of $Z$ and $Z'$:
\bea
Z_1  &=&Z\cos\phi - Z'\sin\phi,\nonumber\\
Z_2  &=&Z\sin\phi + Z'\cos\phi,\nonumber \eea
where $\phi$ is a mixing angle.\\

\hs Desirable symmetry breaking
$ \mbox{SU(3)}_L \otimes \mbox{U(1)}_N\rightarrow \mbox{U(1)}_Q$
and fermion mass generation can be achieved
by introducing three $\mbox{SU(3)}_L$ scalar triplets $\Phi$, $\Delta$,
$\Delta'$ and a sextet $\eta$~\cite{dng}
\bea
&\mbox{SU}(3)_{C}&\hspace*{-0.2cm}\otimes \
\mbox{SU}(3)_{L}\otimes
\mbox{U}(1)_{N}\nonumber \\
&\downarrow      &\hspace*{-0.8cm}\langle \Phi \rangle   \nonumber \\
&\mbox{SU}(3)_{C}&\hspace*{-0.2cm}\otimes \
\mbox{SU}(2)_{L}\otimes
\mbox{U}(1)_{Y}\nonumber \\
&\downarrow      &\hspace*{-0.8cm}\langle \Delta \rangle, \langle
\Delta' \rangle, \langle \eta \rangle\nonumber   \\
&\mbox{SU}(3)_{C}&\hspace*{-0.2cm}\otimes \ \mbox{U}(1)_{Q},
\nonumber \label{ssb2} \eea where the scalar
multiplets are expressed as \bea \Phi & = &\left(
\begin{array}{c}
                \phi^{++}\\ \phi^+\\ \phi^{0}\\

                \end{array}  \right) \sim (1, 3, 1),\nonumber \\
\label{mh1} \Delta & =& \left( \begin{array}{c}
                \Delta^+_1\\ \Delta^0\\ \Delta^-_2\\
                \end{array}  \right) \sim (1, 3, 0),\nonumber \\
\Delta' & =& \left( \begin{array}{c}
                \Delta^{'0}\\ \Delta^{'-}\\ \Delta^{'--}\\
                \end{array}  \right) \sim (1, 3, -1),\nonumber \\
\eta & = & \left( \begin{array}{ccc}
\eta^{++}_1 & \eta^+_1/ \sqrt{2} & \eta^0/ \sqrt{2}\\
\eta^+_1/ \sqrt{2} & \eta^{'0} & \eta^-_2/ \sqrt{2}\\
\eta^0/ \sqrt{2} & \eta^-_2/ \sqrt{2} & \eta^{--}_2
                \end{array}  \right) \sim (1, 6, 0).\nonumber
\eea\\
\hs The sextet $\eta$ is necessary to give masses to
charged leptons~\cite{dng}. The vacuum expectation value
(VEV) $\langle \Phi^T \rangle = ( 0, 0, u/ \sqrt{2})$
yields masses for the exotic quarks, the heavy
gauge bosons $Z_2$ and $X^{\pm \pm},
Y^{\pm}$. The masses of the standard gauge bosons and the ordinary
fermions are related to the VEVs of the other scalar fields,
$\langle \Delta^0 \rangle = v/ \sqrt{2}, \langle \Delta'^0
\rangle = v'/ \sqrt{2}$ and $\langle \eta^0 \rangle = \omega/
\sqrt{2},\ \langle \eta^{'0} \rangle =0 $.
 In order to be consistent with the low energy phenomenology
the mass scale of  $\mbox{SU(3)}_L \otimes \mbox{U(1)}_N$
breaking has to be much larger than the electro-weak scale, i.e.,
$u \gg\ v,\ v',\ \omega$. The masses of
gauge bosons are explicitly: \be
m^2_W=\frac{1}{4}g^2(v^2+v^{'2}+\omega^2),\
M^2_Y=\frac{1}{4}g^2(u^2+v^2+\omega^2),
M^2_X=\frac{1}{4}g^2(u^2+v^{'2}+4 \omega^2), \label{mnhb} \ee and
\bea m_{Z}^2   &=&\frac{g^2}{4 c_W^2}(v^2+v^{'2}+\omega^2)=
\frac{m_W^2}{c_W^2},\nonumber \\
M_{Z'}^2 &=&\frac{g^2}{3}\left[\frac{c^2_W}{1 - 4 s^2_W} u^2 +
\frac{1 - 4 s^2_W}{4 c^2_W}( v^2 + v^{'0} + \omega^2 ) + \frac{3
s^2_W}{1 - 4 s^2_W} v^{'2}\right]. \label{masmat}
\eea\\

Expressions in (\ref{mnhb}) yields a splitting between the
bileptons masses~\cite{lng} \be | M_X^2 - M_Y^2 | \leq 3\  m_W^2.
\label{maship} \ee

\hs By matching the gauge coupling constants we get a relation
between $g$ and $g_N$ -- the coupling constants associated with
$\mbox{SU(3)}_L$ and $\mbox{U(1)}_N$, respectively: \be
\frac{g_N^2}{g^2} = \frac{6 \ s^2_W(M_{Z_2})}{1 - 4
s^2_W(M_{Z_2})}, \label{coupm}
\ee\\
where $e = g\  s_W$ is the same as in the SM.\\

\hs We now discuss  the  previous empirical estimates of the new
gauge boson masses. Combining constraints from direct searches and
neutral currents, one obtains a bound for the mixing angle
 $\phi$~\cite{dng}
and a lower bound on the mass of  $M_{Z_2}$ as $-1.6 \times 10^{-2}
 \le \phi \le 7 \times 10^{-4}$ and
 $M_{Z_2}\ge 1.3 $ TeV, respectively. Such a small mixing angle can
safely be neglected. Adding the constraints from ``wrong'' muon
decay experiments one obtains a bound  for the mass of $Y^+$ as
 $ M_{Y^+} \ge 230 $~ GeV. By computing the
oblique parameters $S$ and $T$, a lower bound of 370 GeV
for $Y^+$ is derived~\cite{fraha}.
Combining this with the mass splitting given in (\ref{maship}) one
obtain a lower bound around 340 GeV for the mass of $X^{++}$.
With  the new atomic parity violation in cesium, one gets a lower bound for
the $Z_2$ mass~\cite{ltrun}: $M_{Z_2} > 1.2 $ TeV. From the symmetry
breaking it follows that the masses of the new charged gauge
bosons $Y^\pm, \ X^{\pm\pm}$ are less than a half of $M_{Z_2}$,
the allowed decay $Z_2 \rightarrow X^{++}\ X^{--}$ with
$X^{\pm\pm} \rightarrow 2 l^\pm$ provides a unique signature in
future colliders. Considering these results it is not inconceivable
that the 3 - 3 - 1 models manifests itself already at
the scale of ${\cal O}(1) $TeV.\\

\section{Single bilepton production in $e^-\  \gamma $ collisions }

\hs In this paper we are interested in the production of new gauge bosons
which exist in the 3 - 3 - 1 model in future high energy colliders.
We focus on the single bilepton production in the electron-photon
collision:

\be  \gamma (p_1,\lambda ) +  e^-(p_2,\lambda')  \rightarrow
X^{--}(k_2,\tau) \ + \ e^+(k_1,\tau '), \label{p1} \label{pro1}
\ee

\be  \gamma (p_1,\lambda ) +  e^-(p_2,\lambda')  \rightarrow
Y^{-}(k_2,\tau) \ + \ \tilde{\nu}_e(k_1,\tau '). \label{p2}
\label{pro2} \ee

Here $p_i$, $k_i$ stand for
the momenta and  $\lambda$, $\lambda'$, $\tau$, $\tau'$  the  helicities of
the particles, respectively.
\subsection{Production of doubly charged bilepton }

\hs First, we cosider the production  of the doubly charged bilepton
(10). The amplitude for this process can be written as
\begin{equation}
M^{X^{--}}=\frac{ieg}{\sqrt{2}}\epsilon_{\mu}^*(k_2)\epsilon_{\nu}(p_1)
\overline{v}(k_1)A^{\mu \nu}u(p_2),
\end{equation}
where $\epsilon_{\nu}(p_1)$, $\epsilon_{\mu}^{*}(k_2)$ are the
polarization vectors of the photon
$\gamma$ and the $X^{--}$ boson, respectively.
At the tree level, there are three
 Feynman diagrams contributing to the reaction (10), representing the
 {\it s, u, t} - channel exchange depicted in Figure 1.
\vspace*{0.3cm}

\begin{center}
\begin{picture}(400,50)(0,0)
\Photon(-10,35)(5,10){2}{4} \ArrowLine(-10,-15)(5,10)
\ArrowLine(5,10)(55,10) \Photon(55,10)(70,35){2}{4}
\ArrowLine(70,-15)(55,10) \Text(32,18)[]{$e^{-} $}
\Text(-16,40)[]{$\gamma$} \Text(-16,-20)[]{$e^-$}
\Text(80,40)[]{$X^{--}$} \Text(80,-20)[]{$e^+$} \Text(185,-42)[]{
Figure 1: Feynman diagrams for $\ e^- \gamma \rightarrow  X^{--} \
e^+ $}

\Photon(150,35)(200,10){2}{8} \Photon(150,10)(200,35){2}{8}
\ArrowLine(135,-15)(150,10) \ArrowLine(200,10)(150,10)
\ArrowLine(215,-15)(200,10)

\Text(180,5)[]{$e^{-} $} \Text(145,40)[]{$\gamma$}
\Text(130,-20)[]{$e^-$} \Text(220,40)[]{$X^{--}$}
\Text(215,-20)[]{$e^+$}
\Photon(270,35)(310,35){2}{6} \Photon(310,-15)(310,35){2}{8}
\Photon(350,35)(310,35){2}{4} \ArrowLine(270,-15)(310,-15)
\ArrowLine(350,-15)(310,-15)

\Text(330,10)[]{$X^{++} $} \Text(265,40)[]{$\gamma$}
\Text(265,-20)[]{$e^-$} \Text(365,40)[]{$X^{--}$}
\Text(360,-20)[]{$e^+$}
\end{picture}
\end{center}
\vspace*{1.5cm}

\hs The $A^{\mu \nu}$ for the three diagrams are given by
\begin{equation}
A^{\mu\nu}_{s}=\gamma^{\mu}\gamma^5 \frac{q\!\!\!/_s
}{q_s^2}\gamma^{\nu},
\end{equation}
\begin{equation}
A^{\mu\nu}_{u}=\gamma^{\mu}\frac{q\!\!\!/_u
}{q_u^2}\gamma^{\nu}\gamma^5,
\end{equation}
\begin{equation}
A^{\mu\nu}_{t} = \frac{2}{(q^2_t- m^2_X )}\gamma_{\alpha}\gamma^5
[(k_2-q_t)^{\nu}g^{\mu\alpha}+(q_t+p_1)^{\mu}g^{\alpha\nu}-(p_1+
k_2)^{\alpha}g^{\nu\mu}].
\end{equation}
Here,  $q_s=p_1+p_2=k_1+k_2$, $q_u=p_1-k_1=k_2-p_2$,
$q_t=p_1-k_2=k_1-p_2$, and $s = (p_1 + p_2)^2$ is the square of
the collision energy. We  work in the center-of-mass frame and
denote the scattering angle (the angle  between
momenta of the initial electron and the final positron)
by  $\theta $. \\
\hs We have evaluated the $\theta$ dependence of the differential
cross-section $d\sigma/d \cos \theta$ and the energy and the
bilepton mass dependence of the total cross-section $\sigma$.\\
\hs i) We show in fig.3 the behaviour of $d\sigma/d \cos \theta$ at
fixed energy. We have chosen a relatively low value of the bilepton mass
$M_X = 350$ GeV. The differential
cross-section  in fig.3 is at  $\sqrt{s}= 1000$ GeV,
but the behaviour is similar at other values of  $\sqrt{s}$.
Note that $d\sigma/d \cos \theta$ is peaked in the backward and
forward direction. This is due to the $e^-$ pole term in the $u$ - channel
and the  $X^{++}$ pole term in the $t$ - channel, respectively, in
eqs. (14) and (15).\\
\hs ii) The energy dependence of the cross-sections
is shown in fig.4. The same value of the bilepton mass as in i),
$M_X = 350$ GeV, is chosen. The energy range is
$600$ GeV $\leqslant \sqrt{s} \leqslant 2500$ GeV. The curve (a)
is the total cross-section $\sigma$, the curves (b), (c), and (d)
representing cross-sections of the $s$, $t$ and $u$ - channels only,
respectively. We see from the figure that the curve (a) goes through
the minimum value and then it increases smoothly with $\sqrt{s}$.
 At the minimum value, we get $\sigma \simeq 18.3$ pb, which is 
large enough
 to measure the $X^{++}$ production. At TESLA based $e \gamma$ colliders,
 with $\sqrt{s} = 911$ GeV~\cite{rwa} the
measurable value of cross-section is $\sigma \geq 18.8 $ pb.
We emphasize that at high energies the $s$, $t$ -channels give main
contributions to the total cross -section.  This is due to
$\sigma_s \sim \frac{1}{m^2}$,
$\sigma_t \sim \frac{1}{m^2}\times Log(s/m^2)$, while $\sigma_u
\sim \frac{1}{S} $.\\
\hs iii) The bilepton mass dependence of the cross-section
for fixed energy typically, $\sqrt{s} = 2000$ GeV, is shown in fig.5.
The mass range is
$300$ GeV $\leqslant M_X \leqslant 700$ GeV. The total 
cross-section decreases
rapidly as $M_X$ increases from 26 pb for  $M_X = 300$ GeV to 4 pb
for $M_X = 700$ GeV.\\

\subsection{Production of singly charged bilepton }
\hs Next, we consider the production of singly charged bilepton and
antineutrino. There are two Feynman diagrams contributing to reaction (11),
representing the {\it s, t} - channel exchange shown in Fig.2 \vspace*{0.2cm}
\begin{center}
\begin{picture}(400,50)(0,0)
\Photon(-10,35)(5,10){2}{4} \ArrowLine(-10,-15)(5,10)
\ArrowLine(5,10)(55,10) \Photon(55,10)(70,35){2}{4}
\ArrowLine(70,-15)(55,10) \Text(-15,-18)[]{$e^{-} $}
\Text(-15,35)[]{$\gamma$} \Text(30,18)[]{$e^-$}
\Text(80,35)[]{$Y^-$} \Text(80,-15)[]{$\widetilde{\nu_e}$}
\Text(185,-42)[]{ Figure 2: Feynman diagram for $ e^- \gamma
\rightarrow  Y^- \widetilde{\nu_e}$}

\ArrowLine(150,-15)(190,-15) \ArrowLine(230,-15)(190,-15)
\Photon(150,35)(190,35){2}{6} \Photon(190,35)(230,35){2}{6}
\Photon(190,-15)(190,35){2}{6} \Text(200,10)[]{$Y^+ $}
\Text(240,-15)[]{$\widetilde{\nu_e}$} \Text(145,-15)[]{$e^-$}
\Text(240,35)[]{$Y^-$} \Text(145,35)[]{$\gamma$}

\end{picture}
\end{center}
\vspace*{1.5cm}

\hs  The corresponding amplitude $ A^{\mu \nu}$ is given
\begin{equation}
A^{\mu
\nu}_s=\gamma^{\mu}(1-\gamma^5)\frac{q\!\!\!/_s}{q_s^2}\gamma^{\nu},
\end{equation}
\begin{equation}
A^{\mu \nu}_t = \frac{1}{(q_t^2-m_Y^2)}\gamma_{\alpha}(1-\gamma^5)
[(k_2-q_t)^{\nu}g^{\mu\alpha}+(q_t+p_1)^{\mu}
g^{\alpha\nu}-(p_1+k_2)^{\alpha}g^{\nu\mu}].
\end{equation}

The notations used here are the same as in the previous case.
We give some estimates of the cross-section as follows\\
\hs i) In fig.6 we plot $d\sigma/d \cos \theta$
as a function of $\cos \theta$
for the collision energy $\sqrt{s}= 1000$ GeV
and the bilepton mass $M_Y = 300$ GeV. It is shown that
 $d\sigma/d \cos \theta$ is peaked in the forward direction,
 due to the $Y^+$ pole term in the $t$ - channel, but it is flat in the
backward direction.

\hs ii) Fig.7 shows the dependence of the total cross section
$\sigma$ as a function of $\sqrt{s}$. The energy range is
$600$ GeV $\leqslant \sqrt{s} \leqslant 2500$ GeV.
The curve (a) is the total cross-section $\sigma$, the curves
(b) and (c) representing cross-sections of the s and
$t$ - channels, respectively.
We can see from fig.7,
if the bilepton mass is the same as fig.6, $M_Y = 300$
GeV, then the $\sigma$ varies  smoothly
increase  from 6 pb to 15.8 pb, which are easily measurable in the
future colliders.  At TESLA based $e \gamma$ colliders we get $\sigma \geq 9 $ pb.\\
\hs iii) The bilepton mass dependence of the cross-section
at the energy $\sqrt{s} = 2000$ GeV is shown in fig.8. Similarly to fig.5,
 the total cross-section decreases as $M_Y$ increases.

\section{Conclusion }
\hs In conclusion, we have presented the minimal 3 - 3 - 1 model
and the production of the doubly
charged bilepton $X^{--}$ and the singly charged one $Y^-$ in
the $e^- \gamma$
reaction. We see that with the first process, the
differential cross-section
is peaked in both the backward and forward directions, while
the second one, the reaction mainly occurs at 
small scattering angles. Based on the result  it is shown that
if the mass of bilepton is in a range of $300$ GeV, then single bilepton
production in $e^- \gamma$ collisions may give observable value
at moderately high energies. At TESLA and CLIC based $e \gamma$ colliders,
with the intergrated luminosity $L \simeq  9 \times 10^{4} {fb}^{-1} $
one expects  several thousand events.\\
\hs Finally,  we note that the second process
is common for both the minimal version
and the version with right-handed neutrinos, while the first one is
characteristic for the minimal version.\\[1.0cm]
One of the authors (D. V. S.) expresses sincere gratitude to the
Nishina Memorial Foundation for the financial support. He would like to
thank Department of Physics, Chuo University
for the warm hospitality during his visit as a Nishina fellow.
This work is supported in part by Research Grants of Japanese
Ministry of Education and Science (Kiban C) and
Chuo University grant for special research. Financial
support from Swedish International Development Cooperation
Agency (SIDA) through the Associateship Scheme of the Abdus Salam
International Centre for Theoretical Physics, Trieste, Italy is
acknowledged (H. N. L.).\\

\newpage
\setcounter{figure}{2}
\begin{figure*}[t]
\centerline{\epsfxsize=12cm\epsffile{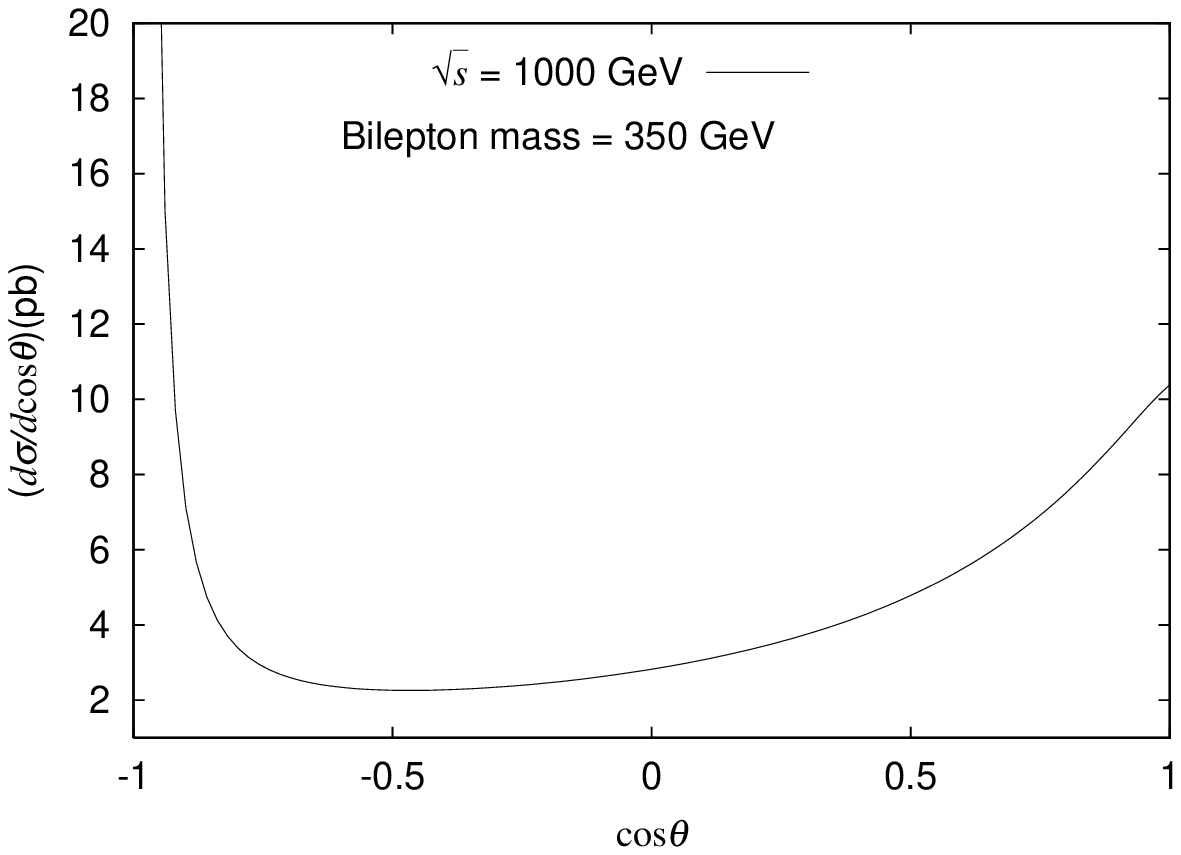}}
\caption{\label{fig3}{\rm Differential cross-section of the
process $e^-
\gamma \rightarrow X^{--} e^+$  as a function of
$\cos\theta$. The collision energy is taken to be
$\sqrt{s}= 1000$ GeV and bilepton mass $M_X = 350$ GeV }}
\end{figure*}
\setcounter{figure}{3}
\begin{figure*}[b]
\centerline{\epsfxsize=12cm\epsffile{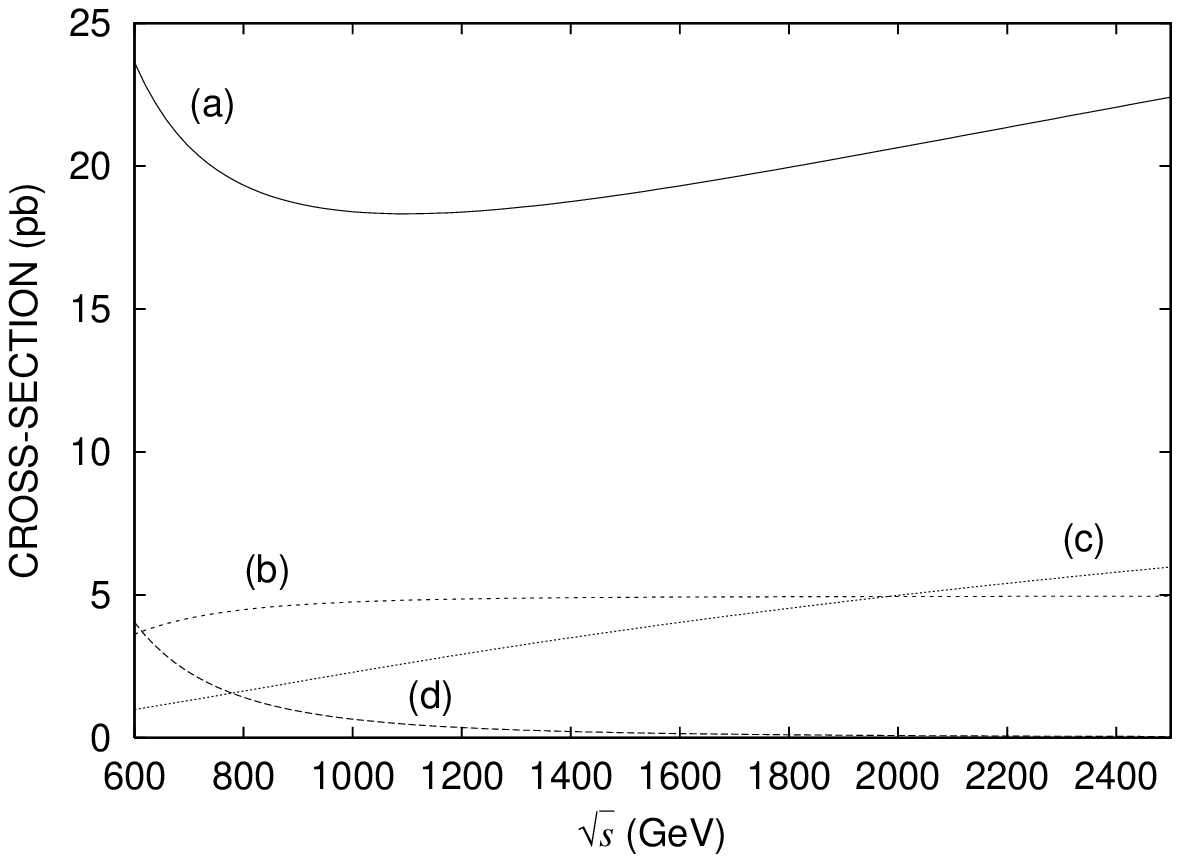}}
\caption{\label{fig4}{\rm Cross-section of the
process $e^- \gamma
\rightarrow X^{--}e^+ $ as a function of
the collision energy  $\sqrt{s}$. The bilepton mass is taken to be
$M_X=350$ GeV. The curve (a)
is the total cross-section $\sigma$, the curves (b), (c), and (d)
representing cross-sections of the $s$, $t$ and $u$ - channels, 
respectively.
}}
\end{figure*}

\setcounter{figure}{4}
\begin{figure*}[t]
\centerline{\epsfxsize=12cm\epsffile{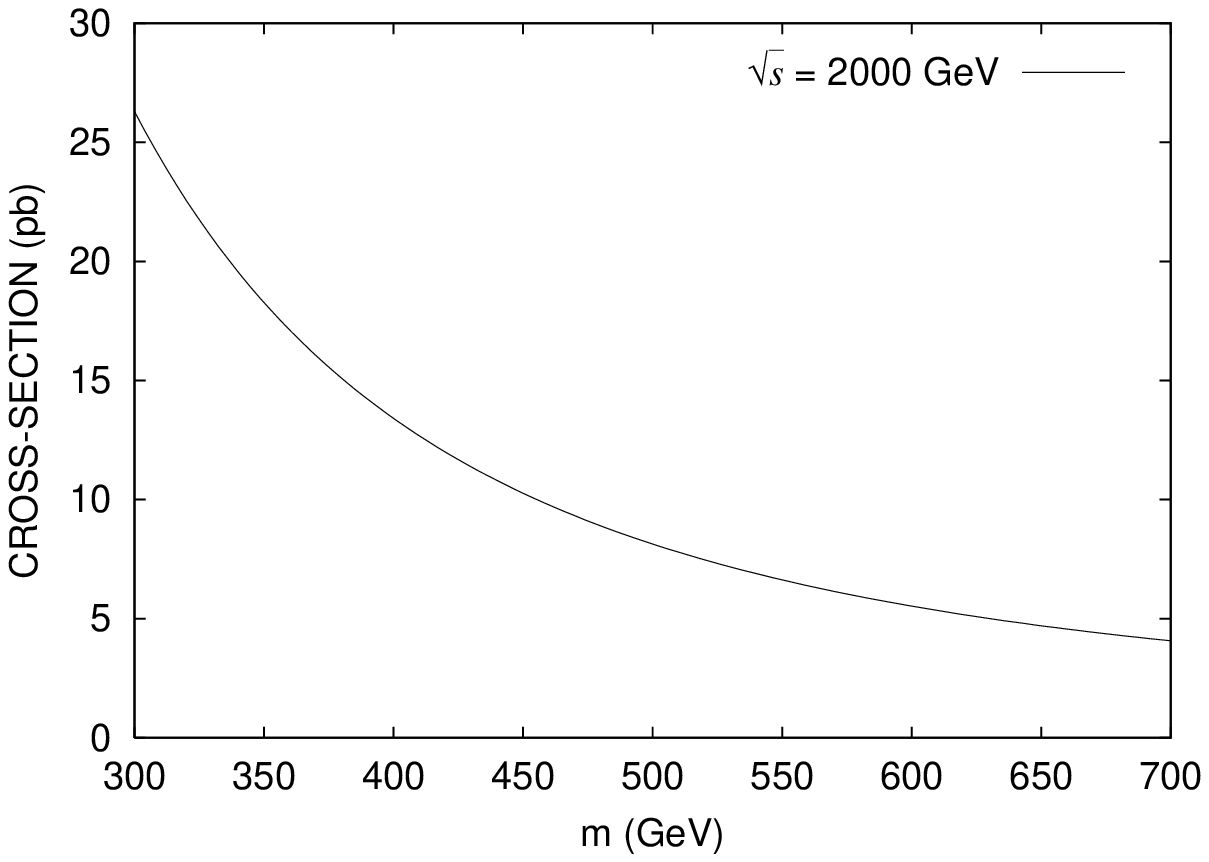}}
\caption{\label{fig5}{\rm  Cross section of the
process $e^- \gamma
\rightarrow X^{--} e^+$ as a function of
 $M_X$. The collision energy is taken to be 2000 GeV.}}
\end{figure*}

\setcounter{figure}{5}
\begin{figure*}[b]
\centerline{\epsfxsize=12cm\epsffile{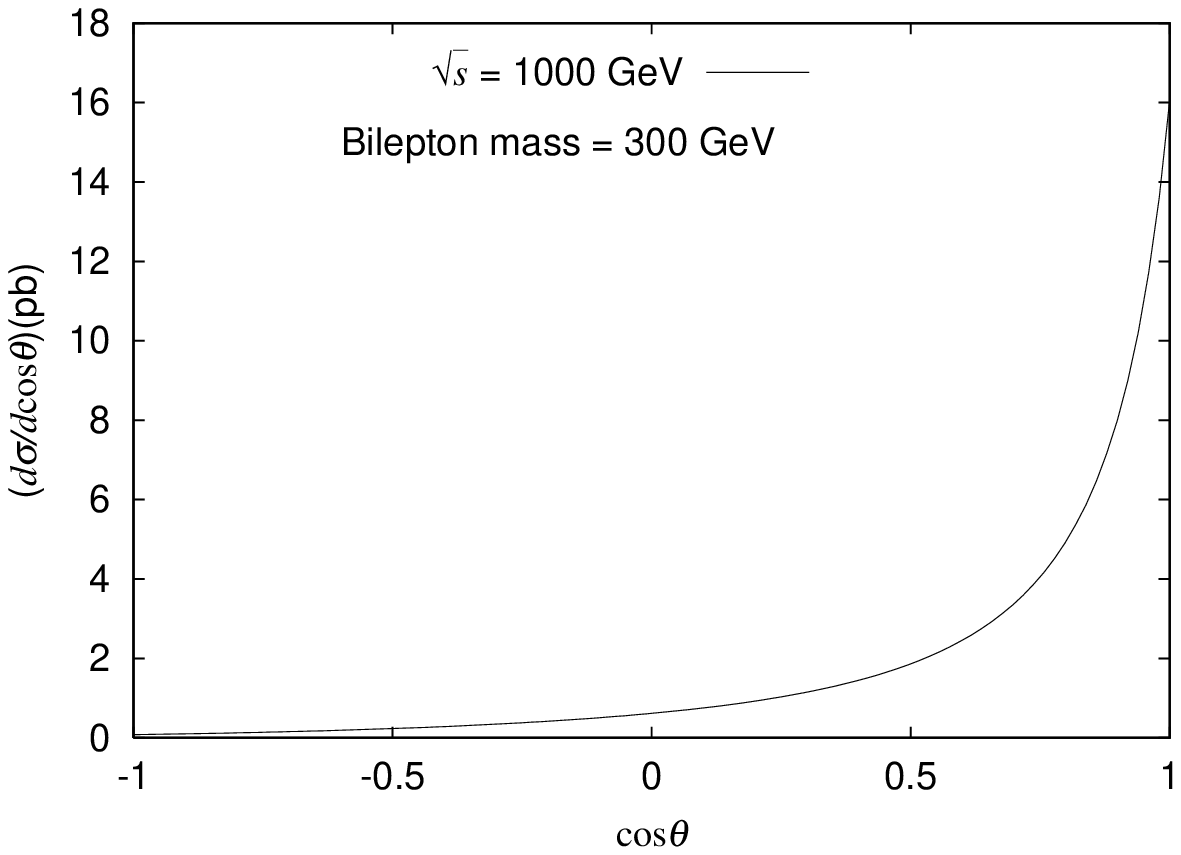}}
\caption{\label{fig6}{\rm Different cross section of the
process $e^-
\gamma \rightarrow Y^- \widetilde{\nu_e}$  as a function of
$cos \theta$, the collision energy $\sqrt{s}= 1000$ GeV.
}}
\end{figure*}

\setcounter{figure}{6}
\begin{figure*}[t]
\centerline{\epsfxsize=12cm\epsffile{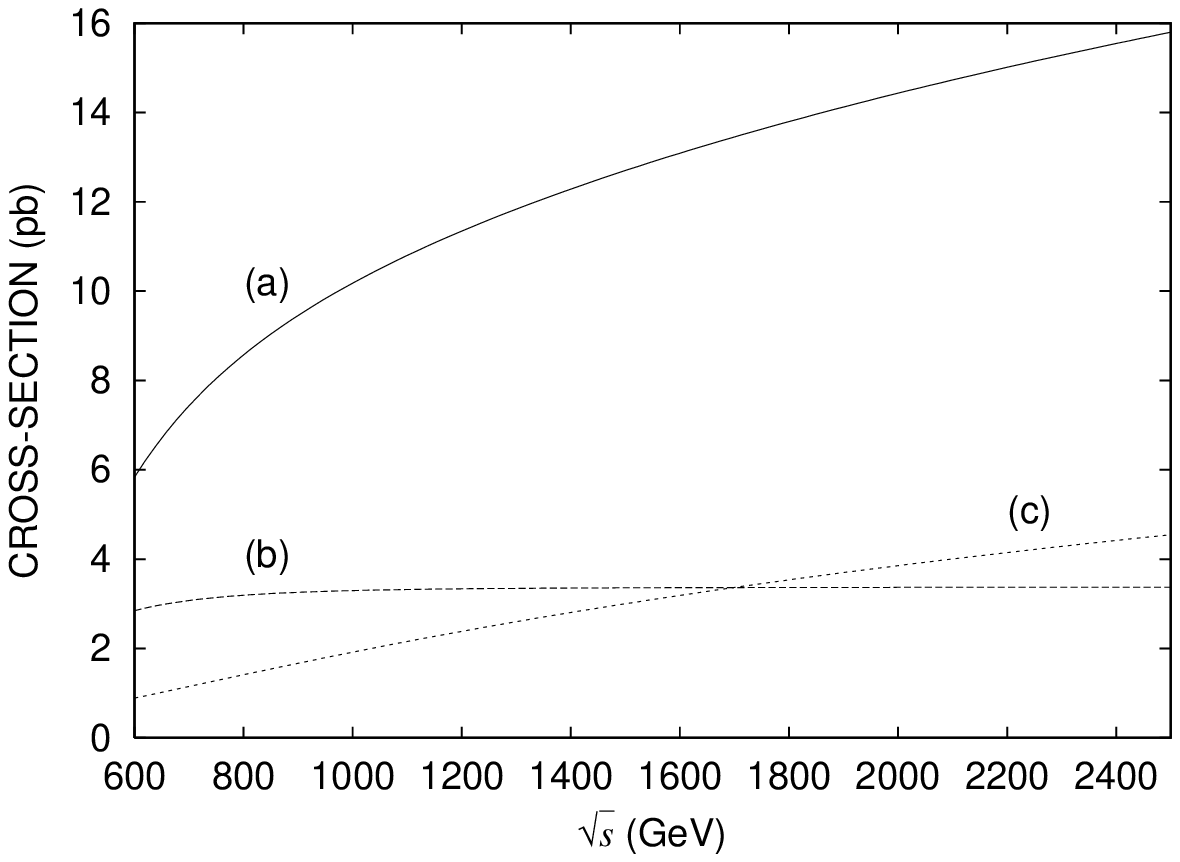}}
\caption{\label{fig7}{\rm Cross section of the
process $e^- \gamma
\rightarrow Y^- \widetilde{\nu_e}$  as a function of
$\sqrt{s}$, bilepton mass is taken to be 300 GeV .
The curve (a) is the total cross-section $\sigma$, the curves
(b) and (c) representing cross-sections of the $s$ and
$t$ - channels, respectively.
}}
\end{figure*}
\setcounter{figure}{7}
\begin{figure*}[b]
\centerline{\epsfxsize=12cm\epsffile{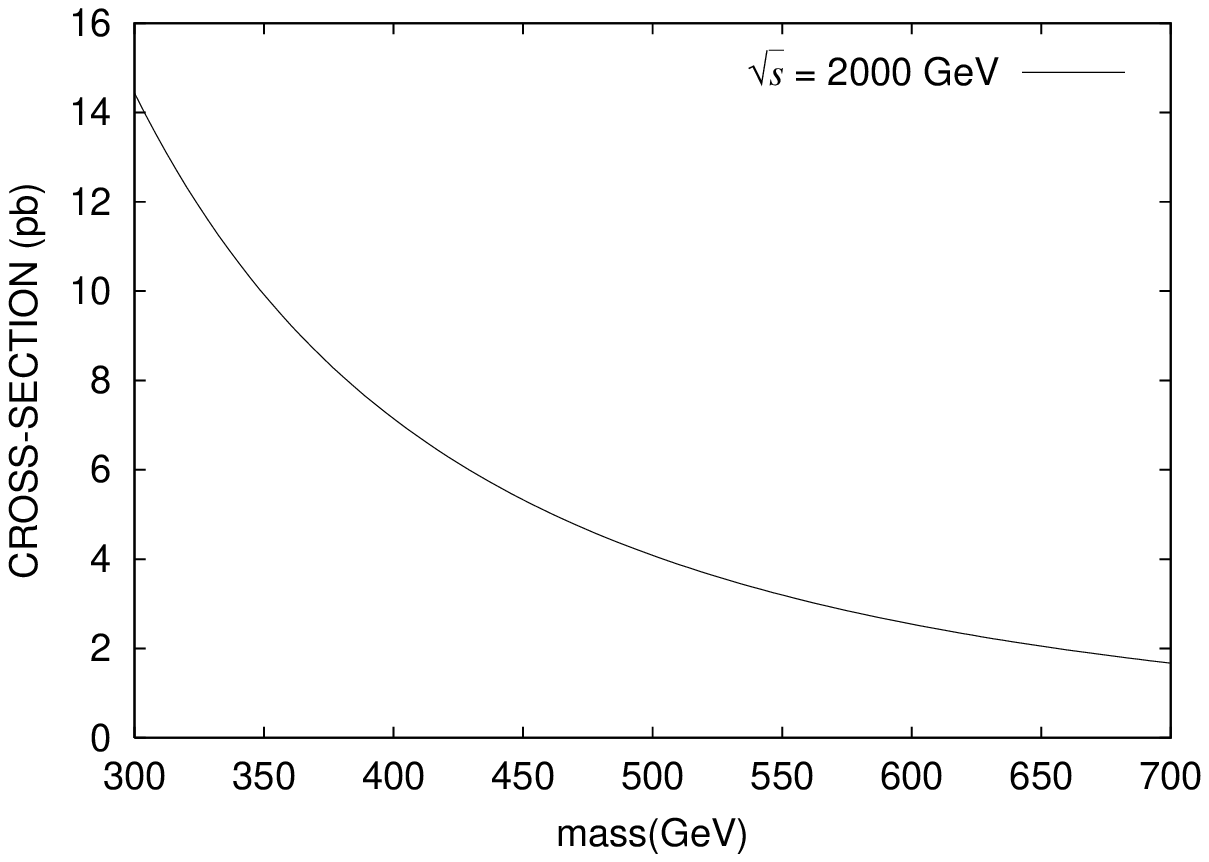}}
\caption{\label{fig8}{\rm Cross section of the
process $e^- \gamma
\rightarrow Y^- \widetilde{\nu_e}$  as a function of $M_Y$.
 The collision energy is taken to be 2000 GeV.
 }}
\end{figure*}
\end{document}